\documentclass[a4]{iopart}  
\usepackage[T1]{fontenc}
\usepackage[latin1]{inputenc}  
\usepackage{hyperref}  
\usepackage{geometry}
\usepackage{graphicx}  
\usepackage{amssymb}  
    
\date{\today}  
\def \be{\begin{equation}}
\def \bea{\begin{eqnarray}}
\def \eea{\end{eqnarray}}
\def \ee{\end{equation}}
\def \no{\nonumber}
\def \a {\alpha}

\def \eps {\epsilon}
\def \r{{\bf r}}
\def \h {\frac{1}{2}}
\def\lsim{\mathrel{\rlap{\lower4pt\hbox{\hskip1pt$\sim$}}
    \raise1pt\hbox{$<$}}}                
\def\gsim{\mathrel{\rlap{\lower4pt\hbox{\hskip1pt$\sim$}}
    \raise1pt\hbox{$>$}}}                

\begin{document}  
\title{Optimising LISA orbits: The projectile solution}  
\author{S. V. Dhurandhar$^1$, K. R. Nayak$^2$ and J-Y. Vinet$^3$}  
\address{ $^1$ IUCAA, Postbag 4, Ganeshkind, Pune - 411 007, India.   
\\$^2$ IISER, Kolkata, India. 
\\ $^3$ Observatoire de la Cote d'Azur, BP 4229, 06304 Nice, France.  
}  
  
\begin{abstract}  
LISA is a joint space mission of the NASA and the ESA for detecting low frequency gravitational waves (GW) in the band 
$10^{-5} - 0.1$ Hz. The proposed mission will use coherent laser beams which will be exchanged between three identical spacecraft forming a giant (almost) equilateral triangle of side $5 \times 10^6$ kilometres. The plane of the triangle will make an angle of  $\sim 60^{\circ}$ with the plane of the ecliptic. The spacecraft constituting LISA will be freely floating in the ambient gravitational field of the Sun and other celestial bodies.  To achieve the requisite sensitivity, the spacecraft formation should remain stable, one requirement being, the distances between spacecraft should remain as constant as possible - that is the flexing of the arms should be minimal. In this paper we present a solution - the projectile solution - which constrains the flexing of the arms to below 5.5 metres/sec in a three year mission period. This solution is obtained in the field of the Sun and Earth only, which principally affect the motion of the spacecraft, especially the flexing of LISA's arms.  
\end{abstract} 

\maketitle  
\section{Introduction \label{SC:1}}  

LISA - Laser Interferometric Space Antenna - is a proposed mission of the ESA and NASA which will use coherent laser beams exchanged between three identical spacecraft forming a giant (almost) equilateral triangle of side $5 \times 10^6$ kilometres for observing low frequency cosmic GW \cite{RIP}. This will complement the ground-based detectors which are geared to operate at higher frequencies ranging from few tens of Hz to kHz.  For the successful operation of LISA it is crucial that the formation of spacecraft be stable - that is, the spacecraft should maintain as much as possible, constant distances between them. However, the spacecraft are freely floating in the ambient gravitational field of the Sun, planets and other celestial bodies (moon for instance) and it is a astrometry problem to seek spacecraft orbits which maintain the equilateral triangular formation as nearly as is possible - that is, optimal orbits for the spacecraft should be found. There are several criteria which the spacecraft formation should satisfy for LISA's successful operation - constraints on, variation in armlengths, the angles  between arms, etc. Here we  focus on the variation in armlengths, the so-called `flexing' of the arms for the reasons detailed below. Optimisation of LISA orbits will be also useful in simplifying the hardware that will be required in the design of LISA. 
\par
Minimising the flexing of the arms is important for  suppressing the laser frequency noise. In ground-based detectors, the near exact symmetry between the arms suppresses this noise as it is common to both arms. But in LISA such high symmetry is not possible, and moreover, the armlengths change with time. Suppression of this noise is crucial since the raw laser noise is orders of magnitude larger than other noises in the interferometer. In LISA, six data streams arise from the exchange of laser beams between the three spacecraft. The  cancellation of the noise is achieved by the technique called time-delay interferometry (TDI) where the six data streams are combined with appropriate time-delays \cite{Arm}. This is possible because of the redundancy present in the data. TDI was put on a sound mathematical footing by establishing that the data combinations had an algebraic structure. The time delayed data is represented by polynomials of time-delay operators acting on the data, each time-delay operator playing the part of an `indeterminate' of a polynomial ring. The data combinations are then represented by polynomial vectors which form a free module over the polynomial ring of time-delay operators. Out of these, the data combinations cancelling laser frequency noise form a submodule of this free module, wellknown in mathematics, as the {\it first module of syzygies} \cite{DNV}. The generators of the submodule were found assuming constant armlengths, where one then deals with the simpler case of a commutative ring of time-delay operators \cite{DNV,NV}. But for realistic spacecraft orbits, the armlengths vary with time, and then the TDI methods involve non-commutative operators leading to the imperfect cancellation of laser frequency noise or the presence of residual noise. The residual noise in turn depends on the rate of change of armlengths - the flexing of arms; thus searching for orbits which reduce the flexing also reduces the residual laser frequency noise.  
\par
 In this paper, following \cite{DVN08}, we include the gravitational field of the Earth in addition to that of the Sun's in the optimisation problem. The orbits in the Sun's field are taken upto second order in $\a$ (or eccentricity), where $\a = l / 2R$, where $l \sim 5 \times 10^6$ km is the nominal distance between the spacecraft and $R$ is one astronomical unit $\sim 1.5 \times 10^8$ km. We find the  perturbative approach for the Sun's field convenient because we also introduce the Earth's effect perturbatively. The second order terms in $\a$ involve the Sun's field upto the octupole order and as shown in \cite{NKDV} almost exactly replicate the Keplerian orbits of the spacecraft and therefore also the flexing. We then linearly superpose the perturbative effect of the Earth's gravitational field over the Sun's field. We choose the Earth over Jupiter because the Earth perturbs the Keplerian orbit in resonance, resulting in unbounded growing of the perturbations and also as shown in \cite{DVN08} Jupiter's tidal field which affects the flexing is less than 10$\%$ of the Earth's and hence not a dominant one. Although we recognise that the problem is inherently non-linear - it is a three body problem - the linear perturbative approach we believe will be useful for short mission periods and also provide directions towards solving the fully general optimisation problem. The analytic approach which we follow here helps to gain insight into the problem. 
\par
We first obtain the general solution  containing 18 arbitrary constants, corresponding to 3 positions and 3 velocities for each of the three spacecraft. Optimising the 18 parameter solution is a daunting problem - we do not attempt to do so here. However, from physical considerations, we present a solution, which we call the `projectile' solution which considerably reduces the flexing of the arms - the rate of change of all armlengths is less than 5.5 metres/sec in a three year mission period. We believe  that these insights will lead us to the full solution of optimisation on 18 parameters. (The nomenclature `projectile' solution will be justified later in the text). 
 
\section{The general perturbative solution of the spacecraft orbits}

  In the subsection below, we briefly summarise the results of the previous papers \cite{DVN08, NKDV} and write down the perturbed CW equations in the two small parameters which describe the effects of the Sun and Earth. In the next subsection we write down the general solution with 18 arbitrary constants; these are the 18 parameters to be varied in order to optimise LISA's orbits with respect to given criteria. Here, our sole criterion is minimising the flexing of the arms.
\par
 Alternatively, we could have chosen to work with the exact Keplerian orbits of the spacecraft for the Sun's field and added to these the perturbations due to the Earth as we have done in \cite{DVN08}. However, here we use the approximate solution to the second order in the  parameter $\a$ because with this solution we gain important physical insights into the problem - for example, we can slightly adjust the tilt of the plane of LISA, by choosing certain constants judiciously, which then helps in  reducing the flexing of the arms. Moreover, we have shown in \cite{NKDV}, we lose very little in accuracy, because the approximate solution is extremely close to the exact. Also it is gratifying to check that, although we end up with 12 arbitrary constants for each spacecraft in this approach, they combine two by two, to yield only six independent arbitrary constants corresponding to the six initial conditions on the three position coordinates and three components of the velocity. 

\subsection{The perturbed Clohessy-Wiltshire equations}

 If we consider only the Sun's field, there exist orbits in which the plane of the LISA triangle makes an angle of about $60^{\circ}$ with the ecliptic and the cluster rolls once per year and for which the armlengths remain constant upto a percent. For these orbits, to the first order in the eccentricity, the distances between spacecraft remain constant; only at the second order in eccentricity the variations in armlengths appear. It was shown in this case that the flexing could be reduced to a minimum  $\sim 48,000$ km \cite{NKDV} by judiciously choosing the orbital parameters of the spacecraft. For establishing this result, it was found convenient to use the Clohessy-Wiltshire equations \cite{CW}. 
\par
Clohessy and Wiltshire make a transformation to a frame - the CW frame $\{x, y, z\}$   which has its origin on the reference orbit and also rotates with angular velocity $\Omega$. The $x$ direction is normal and coplanar with the reference orbit, the $y$ direction is  tangential and comoving, and the $z$ direction is chosen orthogonal to the orbital plane. They write down the linearised dynamical equations for test-particles in the neighbourhood of a reference particle (such as the Earth). The length scale here is the Earth-Sun distance of 1 A. U. and the motion of a test particle is described by these equations if its distance from the origin is small compared with this length scale. Since the frame is noninertial, Coriolis and centrifugal forces appear in addition to the tidal forces.     
\par
We take the reference particle to be orbiting in a circle of radius $R$ with constant angular velocity 
$\Omega$. Then the transformation to the  CW frame $\{ x, y, z \}$ from the barycentric frame $\{ X, Y, Z \}$ is given by,  
\begin{eqnarray}  
x & = & \left(X-R\,\cos\Omega t\right)\,\cos\Omega t\;+\;\left(Y-R\,\sin\Omega t\right)\,  
\sin\Omega t\,,\nonumber \\  
y & = & -\left(X-R\,\cos\Omega t\right)\,\sin\Omega t\;+\;\left(Y-R\,\sin\Omega t\right)\,  
\cos\Omega t\,,\nonumber \\  
z & = & Z.
\label{eq:CW}  
\end{eqnarray}  
The unperturbed CW equations for a test particle with coordinates $(x, y, z)$ are given by,  
\bea  
\ddot{x}-2 \Omega \dot{y} - 3 \Omega^2 x & =& 0 \,  
 , \no \\  
\ddot{y} + 2 \Omega \dot{x} & = &0 \, , \no\\  
\ddot{z} +  \Omega^2 z & = &0.  
\label{gde2}  
\eea  
These equations include terms upto the quadrupole, when the Sun's field is Taylor exanded about the origin of the CW frame. 
The solutions to these equations we call the zero'th order. Among these we choose the solutions which form an equilateral triangular configuration of side $l$. For the $k$th spacecraft, $k = 1, 2, 3$ we have the  following coordinates:
\bea
x_k &=&-\frac{1}{2} \rho_0 \cos(\Omega t - 2 \pi (k - 1) /3 - \phi_0) \, , \no \\
y_k &=& \rho_0 \sin( \Omega t - 2 \pi (k - 1) /3 - \phi_0) \, , \no \\
z_k &=& -\frac{\sqrt{3}}{2}\rho_0 \cos(\Omega t - 2 \pi (k - 1) /3 - \phi_0) \, ,
\label{cws}
\eea 
where $\rho_0 = l / \sqrt{3}$ is the constant distance each spacecraft maintains from the origin of the CW frame and $\phi_0$ is an arbitrary constant phase. In this solution, any pair of spacecraft maintain the constant distance $l$ between each other.
\par
In \cite{NKDV} we have shown that if we include the octupolar terms and solve perturbatively using the zeroth order solution as given by Eq.(\ref{cws}), we obtain the flexing of the arms due to the Sun's field only. We now include the Earth's field as well. LISA follows the Earth $20^{\circ}$ behind. We consider the model where the centre of the Earth leads the origin of the CW frame by $20^{\circ}$ - thus in our model, the `Earth' or the centre of force representing the Earth, follows the circular reference orbit of radius 1 A. U. Also the Earth is at a fixed position vector $\r_{\oplus} = (x_{\oplus}, y_{\oplus}, z_{\oplus})$ in the CW frame. We find that $x_{\oplus} = - R (1 - \cos 20^{\circ}) \sim - 9 \times 10^6$ km, $y_{\oplus} = R \sin 20^{\circ} \sim 5.13 \times 10^7$ km and $z_{\oplus} = 0$. In order to write the CW equations in a convenient form we first define the small parameter $\eps$ in terms of the quantity $\omega_{\oplus}^2 = G M_{\oplus} / d_{\oplus}^3$, where 
$d_{\oplus} = |\r_{\oplus}| \sim 5.2 \times 10^7$ km is the distance of the Earth from the origin of the CW frame; we define 
$\eps = \omega_{\oplus}^2 /  \Omega^2 \sim 7.16 \times 10^{-5}$ which is essentially the ratio of the tidal force exerted by the Earth to that of the Sun. We approximate  $|\r - \r_{\oplus}|$ by $d_{\oplus}$ in the force field of the Earth. We then linearly add the two perturbative terms, namely, the terms describing the  octupolar field of Sun  and the Earth's field and obtain the perturbed CW equations:
\bea  
\ddot{x}-2 \Omega \dot{y} - 3 \Omega^2 x +  \frac{3 \a \Omega^2}{l}(2 x^2 - y^2 -z^2) + \eps \Omega^2 (x - x_{\oplus}) & =&  0 \,, \no \\  
\ddot{y} + 2 \Omega \dot{x} - \frac{6 \a \Omega^2}{l} x y + \eps \Omega^2 (y - y_{\oplus}) & = & 0 \,, \no\\  
\ddot{z} +  \Omega^2 z  - \frac{6 \a \Omega^2}{l} x z + \eps \Omega^2 z & = &0.  
\label{CWcomb}  
\eea
We now have the perturbed equations in two small parameters $\a$ and $\eps$.  We  seek perturbative solutions to Eq. (\ref{CWcomb}) to the first order in $\a$ and $\eps$. We note that the forcing terms by the Earth  in these equations appear at the same frequency $\Omega$ and hence they imply resonance. This means that the Earth's effect on LISA is cumulative and therefore important. Also, we have here ignored higher order terms in both $\a$ and $\eps$ as well as the cross terms in these parameters, in order that the problem becomes linear and therefore tractable. Thus the solutions we will obtain are valid in the short term or for short periods of the LISA mission; for longer periods the problem is inherently nonlinear and difficult to deal with analytically - it is infact a three body problem that we are approximating. 
\par
The quantity $\omega_{\oplus}^{-1} \sim 18.8$ years defines a timescale. If we assume a stationary Earth and Earth's gravitational field only, the free fall time of a particle initially at rest at a distance $d_{\oplus}$ from Earth is about 21 years which is comparable to this timescale. The three year mission period we have assumed here is smaller than the above timescales and therefore sufficiently short for our analysis to be useful. Moreover, as it will turn out for the solution we present, the LISA spacecraft fall towards the Earth about half a million km from their initial positions, thus remaining well within the CW frame. Thus we expect  the linear perturbative analysis that we have carried out here to hold good.  

\subsection{The general solution with 18 parameters (arbitrary constants)}

 Since these solutions have been derived in previous papers \cite{DVN08, NKDV}, we merely state the results here. We adopt the following notation: we denote the solution for spacecraft $k$ with $k = 1, 2, 3$ by the bracketed suffix $k$ and the zeroth order by the suffix $0$, the $\a$ perturbation by the suffix $1$ and Earth's perturbation - the $\eps$ perturbation - by the suffix $2$; thus we write:
\be
x_{(k)} = x_{(k) 0} + \a x_{(k) 1} + \eps x_{(k) 2} \,,
\label{full} 
\ee    
and similarly for the $y$ and $z$ coordinates. These are the coordinates of the spacecraft in the CW frame. Further, to reduce the clutter, we choose units of time and length such that $\Omega = 1$ and $l = 1$. In these units, to the zeroth order, the spacecraft form an equaliteral triangle of side unity with the distance of each spacecraft from the origin equal to $1/\sqrt{3}$; also one year period in these units equals $t = 2 \pi$. In these units we may rewrite Eq.(\ref{cws}) as follows:
\bea
x_{(k) 0} & = & -\frac{1}{2\sqrt{3}} \cos \phi_k  \,, \no \\
y_{(k) 0} & = & \frac{1}{\sqrt{3}} \sin \phi_k \,, \no \\
z_{(k) 0} & = & -\frac{1}{2} \cos \phi_k \,,
\label{zeroth}
\eea
where, $\phi_k = t -  2 \pi (k -1) / 3 - t_0$ and $t_0$ is a arbitrary constant phase.  The perturbative solutions are the following:
\bea
x_{(k) 1} & = & 2 A_k + B_k\cos \phi_k + C_k \sin \phi_k + \frac{5}{8} - \frac{1}{24} \cos 2 \phi_k \,, \no \\
y_{(k) 1} & = & -\left (3A_k + 5/4 \right) t + 2 \left (C_k \cos \phi_k - B_k \sin \phi_k \right) + D_k + \frac{1}{6} \sin 2 \phi_k  \,, \no \\
z_{(k) 1} & = & E_k \cos \phi_k + F_k \sin \phi_k + \frac{\sqrt{3}}{4} - \frac{1}{4\sqrt{3}}\cos 2 \phi_k \,;
\label{alpha}
\eea
and,
 \bea
x_{(k) 2} & = & 2A'_k+x_{\oplus}+2 t y_{\oplus}+B'_k \cos \phi_k + C'_k \sin \phi_k + \frac{5t}{4\sqrt{3}} \sin \phi_k \,,  \no \\
y_{(k) 2} & = &  - 3 A' t - 2 x_{\oplus} t - \frac{3}{2} y_{\oplus}t^{2} +  \frac{5t}{2\sqrt{3}} \cos \phi_k  - 
\frac{\sqrt{3}}{2} \sin \phi_k  \no \\
&& + 2\left(C'_k \cos \phi_k - B'_k \sin \phi_k \right) + D'_k \,, \no \\
z_{(k) 2} & = & E'_k \cos \phi_k + F'_k \sin \phi_k + \frac{1}{4} t \sin \phi_k \,.
\label{epsilon}
\eea
The quantities $A_k, B_k, C_k, D_k, E_k, F_k$ and $A'_k, B'_k, C'_k, D'_k, E'_k, F'_k$ are arbitrary constants. For each spacecraft, there seem to be 12 arbitrary constants. However, if we now add up all the solutions given in Eqs.(\ref{zeroth}), (\ref{alpha}) and (\ref{epsilon}) to obtain the full solutions as in Eq.(\ref{full}), the arbitrary constants combine as 
$\a A_k + \eps A'_k, \a B_k + \eps B'_k, ...$ etc.  to give just six independent arbitrary constants for each spacecraft as demanded by the three second order simultaneous differential equations. We find however, that it is better to leave the arbitrary constants as they are, because from our previous experience, we know what values the arbitrary constants should take in order that the spacecraft form stable or nearly stable configurations in which the variation in armlengths is acceptably small.  

\section{Stability and reduced flexing of the arms of LISA: the projectile solution}

 We seek solutions that are (i) stable, and (ii) reduce the flexing of the arms. Following \cite{NKDV} we satisfy the first criterion by choosing the following values of the constants:
\be
A_k = - \frac{5}{12},~B_k = \frac{1}{16},~C_k = D_k = 0,~E_k = \frac{\sqrt{3}}{16},~F_k = 0,~~k = 1, 2, 3.
\ee 
This choice of constants ensures that, (i) to the first order in eccentricity (or $\a$), the spacecraft maintain constant distances from the origin, forming an equilateral triangle which makes an angle  of $60^{\circ}$ with the ecliptic; (ii) to the second order in $\a$, the spacecraft do not drift away - the choice of $A_k$ ensures that the secular term proportional to $t$ in the $y_{(k) 1}$ is set to zero; now the angle of the plane of the triangle is not exactly $60^{\circ}$, but very close to it - $5 \a /16 \lsim 0.01$ radians from $60^{\circ}$. If we `switch off' the Earth's field, this choice of tilt angle (of constants) ensures that the flexing is kept at a minimum to $\lsim 48,000$ km which is less than $1 \%$ variation in the armlength. Thus,  even with the Earth's field we seek solutions that are close to the previously found solutions,  which were shown to have optimal properties in the field of the Sun only. With this `safe' strategy, we expect not to stray away from optimality. The solutions that we will find do not lay any claim to exact optimality, but they do exhibit adequate reduced flexing of the arms to $\lsim 5.5$ m/sec in a 3 year mission, which would inturn be useful for reducing the residual laser frequency noise in TDI.  In \cite{DVN08} it has been shown that in the type of solutions we are considering, it is essentially the ${\dot L}$ terms which contribute to the residual noise, where $L$ generically is the length of any one arm; higher order terms can be neglected. Moreover, the amplitude of the residual noise is $\propto {\dot L}$ and thus its power spectral density (PSD) is $\propto {\dot L}^2$. Thus a reduction in ${\dot L}$ from 
10 m/sec say, which was the estimate in earlier literature, to 5.5 m/sec, reduces the PSD of the residual noise to almost 
$30 \%$ of its earlier estimate. 
\par
We now turn to the primed set of constants which occur in the Earth's perturbative part of the solution. In this part we are guided by a physical criterion. We would like the Earth's perturbative effect to be small during LISA's mission. One way to achieve this is by setting $x_{(k) 2} = y_{(k) 2} = z_{(k) 2} = 0$ as also the velocities 
${\dot x}_{(k)2} =  {\dot y}_{(k)2} = {\dot z}_{(k)2} = 0$ at an appropriate epoch $t$, where the overdot represents the time derivative of a quantity. This appropriate epoch we choose at the middle of the mission; so if the mission period is $T$ and we arrange so that the appropriate epoch occurs at $t = 0$, then the mission duration is $- T/2 \leq t \leq  T/2 $. With these initial conditions the primed set of arbitrary constants are determined. Defining for the spacecraft $k = 1, 2, 3$, the constant phases 
$t_k = t_0 + 2 \pi (k - 1) / 3$, and imposing the above mentioned initial conditions at $t = 0$, the arbitrary constants for the spacecraft take the values:
\bea
A'_k &=& - \frac{1}{\sqrt{3}} \cos t_k \,, \no \\
B'_k  &=& \frac{2}{\sqrt{3}} - x_{\oplus} \cos t_k - 2 y_{\oplus} \sin t_k - \frac{\sqrt{3}}{4} \sin^2 t_k   \,, \no \\
C'_k &=& x_{\oplus} \sin t_k - 2 y_{\oplus} \cos t_k - \frac{\sqrt{3}}{4} \sin t_k \cos t_k \,, \no \\
D'_k &=& 4 y_{\oplus} - \frac{4}{\sqrt{3}} \sin t_k  \,, \no \\
E'_k &=& \frac{1}{4} \sin^2 t_k \,, \no \\
F'_k &=& \frac{1}{4} \sin t_k \cos t_k \,. 
\eea
These constants determine the orbits of the spacecraft and therefore the distances between them as a function of the epoch 
$t$. The armlengths $l_{ij}, i, j = 1, 2, 3$ are then computed by the formulae:
\be
l_{ij} = [(x_{(i)} -  x_{(j)})^2 + (y_{(i)} -  y_{(j)})^2 + (z_{(i)} -  z_{(j)})^2]^{\h}  \,. 
\ee
In the figures (\ref{armlen}) and (\ref{armvel}) below, we plot the armlengths $l_{ij}$ and the rate of change of armlengths ${\dot l}_{ij}$ for the phase $t_0 = 0$ for the mission period of 3 years which in our units is $T = 6 \pi$. The armlength variation increases from the previous optimum obtained in the field of the Sun only, from 48,000 km to roughly 60,000 km in the combined field of the Sun and Earth. Also the rate of change of armlengths increases from the maximum of 4 metres/sec for Sun's field only to a maximum of 5.5 metres/sec in the combined field.
\begin{figure}[h]  
\centering  
\includegraphics[width = 0.6\textwidth]{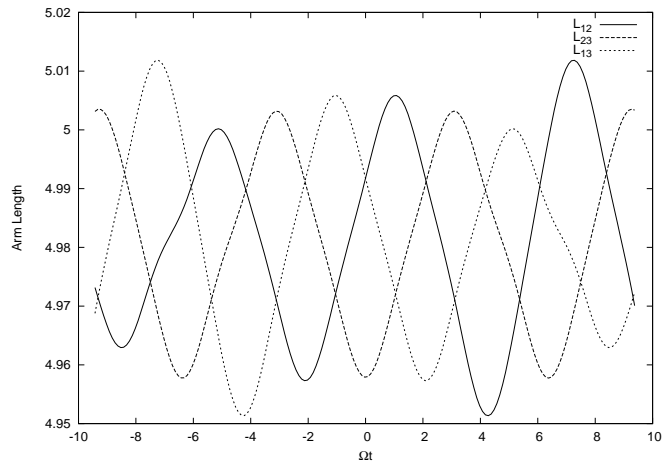}  
\caption{The figure shows the variation in the three armlengths of LISA for a mission period of three  years ($ - 3 \pi \leq \Omega t \leq 3 \pi)$ for the phase $t_0 = 0$ in millions of km. The maximum variation in armlengths is about 60,000 km.}  
\label{armlen}  
\end{figure}  
\begin{figure}[h]    
\centering  
\includegraphics[width = 0.6\textwidth]{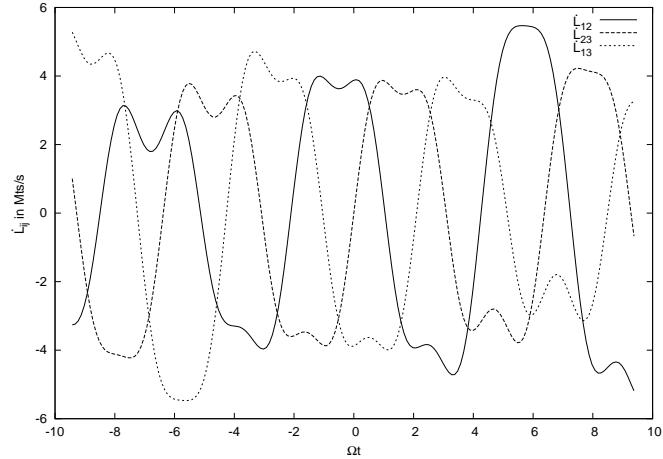}  
\caption{The figure shows the time derivative of the armlengths for a mission period of three years ($ - 3 \pi \leq \Omega t \leq 3 \pi)$ for the phase $t_0 = 0$. The flexing of the arms is less than 5.5 metres/sec.}  
\label{armvel}  
\end{figure}  
\par
\begin{figure}[t]  
\centering  
\includegraphics[width = 0.6\textwidth]{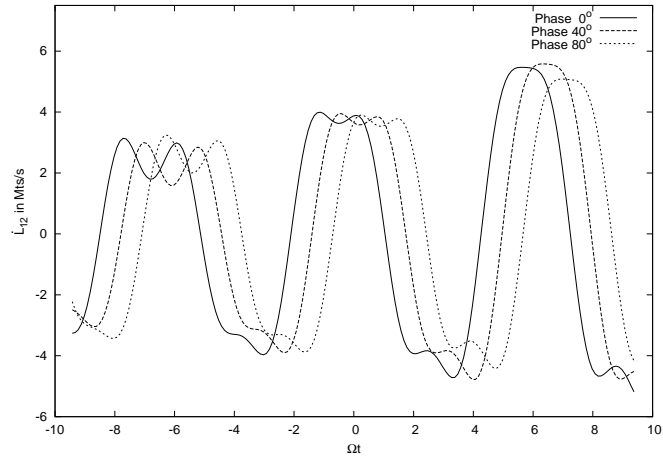}  
\caption{The figure shows the flexing of the arm $l_{12}$ for the different phases $t_0 = 0, 40^{\circ}, 80^{\circ}$ and 
$ - 3 \pi \leq \Omega t \leq 3 \pi$.  The maximum flexing is seen to be insensitive to the phase.}  
\label{armvelphs}  
\end{figure}

We also plot in figure (\ref{armvelphs}) the rate of change of armlength $l_{12}$ for the different phases,  $t_0 =0, 40^{\circ}, 80^{\circ}$. We observe that the profiles of the curves essentially repeat; the curves are basically time translated. The figure shows that the maximum flexing is essentially insensitive to the phase.  This is the consequence of the high symmetry of the LISA configuration.
We term this as the `projectile' solution because just as when a stone (projectile) is vertically thrown from the ground in a gravitational field that is assumed to be constant, the stone reaches zero velocity midway and returns to the ground. The  solution presented here describes an analogous situation;  LISA is `thrown' away from the Earth initially, thus it moves away relative to the Earth in the first half of the mission, reaches maximum distance away from the Earth at mid-mission, and then in the next half of the mission period falls towards the Earth. One can easily compute the distance travelled by LISA relative to the Earth by examining the expressions for $(x_2, y_2, z_2)$  in Eq.(\ref{epsilon}). We have dropped the subscript $k$ in order to avoid clutter - we can do this because of symmetry of the LISA configuration; the results are essentially the same for all spacecraft. Let us therefore consider spacecraft 1. If we consider a 3 year mission period, then $T = 6 \pi$. At initial time and final times, $t = \pm T/2 = \pm 3 \pi$, the coordinate which dominates is the $y$ coordinate (not surprisingly, as this is roughly the direction of the force of the Earth); and the term that dominates in the $y$ coordinate is the quadratic term in $t$. Thus $y_2 (\pm T/2) \sim (3/2 y_{\oplus})( T^2 / 4) = 27 \pi^2 / 2 y_{\oplus} \sim 1367$ in the units chosen, for $T = 6 \pi$. Converting to km by multiplying by the factor $\eps \times 5 \times 10^6$ km yields $\eps y_2 \sim 4.9  \times 10^5$ km. At $t = 0$, the initial condition implies, $y_2 = 0$ and also ${\dot y}_2 = 0$. Thus in this solution, in the first year and half, LISA travels about 500,000 km away from the Earth and then falls back towards the Earth about the same distance in the second half of the mission. It does not fall back exactly to the same point though, even relative to the Earth, because  $x_2 \sim 2 y_{\oplus} t$ for large times and thus $\eps x_2 (\pm 3 \pi) \sim \pm 6 \pi \eps y_{\oplus} \sim \pm 7 \times 10^4$ km.  The $z$ coordinate changes very little and plays a minor role in the solution.
\par
As compared to the solution described in \cite{DVN08}, in which the initial conditions $x_{(k) 2} = y_{(k) 2} = z_{(k) 2} = {\dot x}_{(k)2} =  {\dot y}_{(k)2} = {\dot z}_{(k)2} = 0$ were applied at the start of the mission $t = 0$, and where it was found that the maximum flexing after three years was about 8 metres/sec, here in the projectile solution, the flexing is reduced to $5.5$ metres/sec, in which the same initial conditions are applied mid-mission. This gives an improvement of about $30 \%$  in the maximum flexing, and more than a factor of two in the PSD of the residual laser frequency noise. If this level of residual noise in the TDI observables  can be tolerated, then these spacecraft orbits can be considered to be adequate.   

\section{Concluding Remarks}

 In this paper, we have presented a solution for the LISA spacecraft orbits which gives reduced flexing of LISA's arms to less than 5.5 metres/sec in a three year mission. The solution has been obtained in the combined field of Sun and Earth. The solution although approximate is analytical and hence provides valuable insights into the problem of optimisation. Clearly, this is not the most optimised solution that is possible. The truly optimised solution for short mission periods may be computed by varying the 18 arbitrary constants in the general solution which has been given in this paper. Although we have argued here, that the tidal effects due to Jupiter are small, for a complete solution, it would be desirable to include the field of Jupiter in future endeavours. 
\par
Optimisation of orbits is an important problem for LISA, because the judicious choice of orbits can lead to several advantages. As we have argued here, reducing the flexing of the arms from say 10 metres/sec to 5.5 metres/sec, tends to reduce the PSD of the residual noise in the TDI observables to about $30 \%$ of its original estimated value. This reduction could further help in the simplification of the TDI in which the first generation modified TDI could suffice, thus in turn reducing the degree of the polynomials in the time delay operators. Lower degree polynomials are preferred because they decrease the interpolations required to be carried out on the data and in turn the overall noise. Further, the optimisation of orbits can also help in the simplification of hardware in the design of LISA. 
\ack

The authors S. V. Dhurandhar and J-Y Vinet would like to thank the Indo-French Centre for the Promotion of Advanced Research (IFCPAR) project no. 3504-1 under which this work has been carried out.

\end{document}